# Design Contracts For Networked Automation Systems Co-design

B. Sreram, Seshadhri Srinivasan, B. Subathra
Kalasalingam University
Srivilliputtur, India
seshadhri@ieee.org

Srini Ramaswamy
ABB Inc, USA
srini@ieee.org

*Abstract*— Networked automation systems (NAS) are characterized by confluence of control, computation, communication and Information ($C^3I$) technologies. Design decisions of one domain are affected by the constraints posed by others. Reliable NAS design should address the requirements of the system, and simultaneously meet the constraints posed by other domains and this is called co-design in literature. Co-design requires clear definition of interfaces among these domains. Control design in NAS is affected by the timing imperfections posed by other domains. In this investigation, we first study the different sources of timing imperfections in NAS, and classify them based on their occurrence. The concept of jitter is used to define the timing imperfections induced by various system components. Using this analysis, we classify the jitter based on their behavior and domain of occurrence. Our analysis shows that the jitter induced in NAS can be classified based on domain as- hardware, software and communication. Next, we use this analysis to model the jitter from the components of NAS. Modeling timing imperfections helps in capturing the interfaces among the domains, and we use the concept of design contracts to capture the interfaces. Design contracts describe the semantic mapping among the domains and are specified using the jitter margins. Implementing design contracts requires knowledge of the jitter margin and, the results from control theory are used to this extent.

## I. Introduction

Networked automation systems (NAS) in industrial automation refer to an automation system involving networked sensors, actuators and controllers [1]. NAS have plethora of applications that have tight coordination between physical process and software (see, [70]-[78]). There is a confluence of control, computation, communication and information ($C^3I$) technologies in NAS as shown in Fig. 1. These are domains with little common ground and have conflicting design specifications. For example, feedback control systems have strict timing performance and accuracy of data is not strict, whereas functionality and integrity of the transmitted data are important to computation and communication respectively. As against this industrial automation system must transmit the correct control information using software that achieves the required functionality in a reasonable time over a communication channel. This interplay among the domains and, conflicting requirements alongside assumptions make the design of automation systems complex. Further, design decisions in one domain affect the performance of the other. Such inter-disciplinary dependency and its nature (linear or non-linear, implicit or explicit) affect the performance of NAS. Therefore it becomes necessary to define the interfaces and communication among these conflicting domains. Interfaces need to define the design variables, constraints, and associated trade-offs. In other words, NAS requires capturing the constraints posed by the various domains while meeting the requirement of reliable control in one-go. Such an approach also makes it possible to evaluate design alternatives, and to study different engineering solutions. For example, stability limits of a control system with different communication protocols can be evaluated and the system can be engineered to have better stability limit or a lower stability limit with cost savings. This requires a co-design approach wherein the design requirements are captured to meet the constraints across the various disciplines. This is also important because the domains involved in NAS have-different views, encompassing terminology, theories, techniques, tools and design methods.

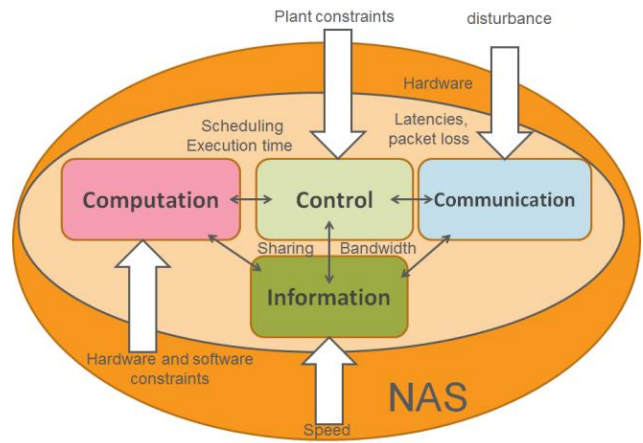

Figure 1. $C^3I$ view of NAS with associated constraints

Functional requirement for NAS is usually defined by the application and timing performance has been identified to be one of the important non-functional requirements in [20-22]. It has been pointed out that the timing performance depends on the application relations and chosen path of communication.

Gaëlle Marsal et al. [23] use the concept of response time which is the time elapsed between the occurrence of an event on the process and the time of occurrence of the response event, to capture the timing specification on NAS. Probabilistic timed automata model is used for verification of timing performance using probabilistic model checking (PMC) in [24]. A simulation approach for evaluating the response time in NAS was studied in [25]. In general, response time cannot capture the time-varying behaviour induced due to imperfections. Stochastic modelling of delays considering three types of signals, namely fixed signal (which maintains its value until changed), cyclic signal (A signal which changes values periodically) and a combination of the two signals has been studied in [26]. But, this analysis cannot encompass all the signals that are occurring in an industrial automation system. Further, it is restricted to a class of signals like logic and sequencing. The use of Markov chains to model delays in communication channels has been recently explored in [27]. The investigation fails to identify the delays induced due to hardware and software, and it also fails to give guidelines or bounds on the control execution. Though timing performance is a non-functional requirement in software engineering it affects the stability and performance of the control systems (see, [28]-[36] and references therein). M. Sanfridson [36] studied the effect of timing performance on real-time systems and identified latency and jitter as the source of imperfections in timing performance. Latencies and jitter in NAS are due to communication and computation imperfections (like network latency and computation jitter). These imperfections affect the stability and reliability of the NAS. Pau Marti et al. [37] studied the various types of jitter that occur in a real-time control system and proposed a control compensation scheme. Further, different scheduling schemes for software were studied for compensating the jitter. It was observed that modifications in scheduling algorithm were not enough to compensate jitter and even small jitters caused degradation in system performance. In [10], theory and method to design controllers considering jitter has been identified as the future direction of research in networked control systems. One can infer from above discussion the importance of co-design considering timing imperfections in NAS. This paper is motivated to this extend.

The main contributions of this investigation are: *(i)* analysis of timing imperfections from various system components in NAS *(ii)* classify the timing imperfections based on their behaviour and domain, *(iii)* model the timing imperfections using suitable tools, *(iv)* propose a co-design approach for NAS to have reliable control in the presence of timing imperfections posed by software and communication using the concept of design contracts, and *(v)* study the implementation of the contract using results from control theory. The results obtained from this investigation are useful for co-designing NAS for reliable control by considering the constraints from other domains (computation, control and hardware), in engineering NAS, benchmarking different automation solutions, and testing and verification of the NAS.

The paper is organized into six sections. In section II, the problem studied is described. Section III studies the timing imperfections in NAS from various system components, and modelling the timing imperfections is discussed in section IV. Timing tolerance contract, its suitability to NAS, and its implementation in the context of NAS is discussed in section V. Conclusions and future directions of this investigation is presented in section V1.

II. PROBLEM DEFINITION

*A. Previous work*

Co-design is used in systems having interplay among various domains, and targeted to meet the system requirements in the presence of complex domain dependencies. Co-design approaches have been used in control systems in the context of networked control systems (see, [2]-[11]), embedded control systems [12-13], hardware, and software [14], electrical and mechanical design [15], embedded systems [16-18] among others. In a recent research survey on co-design approaches for networked control systems- H. Yan et al. [11] commented that the co-design tool should deal with complex dependencies among different domains. Finding rational abstractions that allow (control) system requirements, and assumptions alongside platform constraints to be communicated across the domains was identified as the key issue in such a design. Further, it was also commented that the co-design approach not only leads to optimal design but also gives less conservative results that could be used for design of NCSs. To our knowledge, the use of co-design approach for NAS has not been investigated in the literature. Further, the concept of rational abstractions has not been answered and not many results are available in this direction. Motivated by these gaps, we study NAS co-design for the functional requirement of reliable control in the presence of timing imperfections. We try to answer the questions about suitable rational abstractions in the context of NAS by studying the timing imperfections that affect the control performance. In [19], design contract has been used as tool for defining interface among software and control domain in the context of CPS. The main idea is to extend the idea of design contracts to NAS wherein the contracts specify the timing requirement and, provide interface among the domains.

Contract based design was proposed by B. Meyer [51] in the context of the programming language Eiffel. Contract based design has been used for designing closed loop controllers for hybrid systems [52], embedded systems, Simulink control diagrams [53], robotic systems, heterogeneous systems to mention a few. In all the above cases, contracts have been used as an interface for inter-domain communication and/or negotiation. Recently, the concept of contract based design has been studied in the context of CPS in and various design contracts have been proposed. Further, the investigation also outlines some open problems, they are:

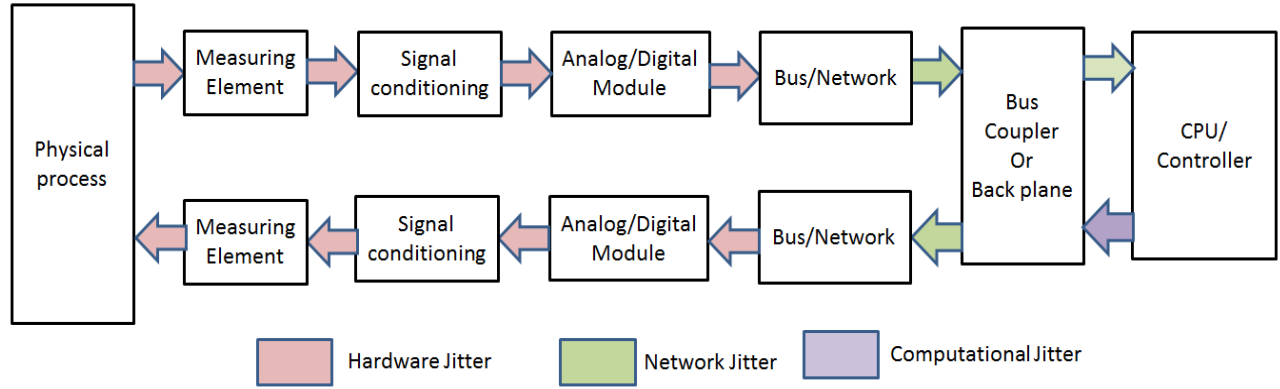

Figure 2. Generic NAS adapted from [41]

1. choosing the contract for a given application (design-space exploration problem)
2. fixing the parameters of the contract (synthesis problem)
3. verify the control design for a given contract (verification problem)
4. deriving software implementation from contracts automatically (a problem suite involving platform-space exploration, mapping, model transformations and compiler optimization)

We try to use the idea of design contract in NAS and find answers to these issues. We use the timing tolerances contract (TOLC) proposed in [19] to specify the contract (i.e. use sampling and delay jitter). Though the results obtained in this investigation are motivated to be used in NAS, they are equally valid for NCSs and CPS as well. One should note here that though design contracts are closely related to design by contract used in software engineering, they are not the same. Design contracts are used to bridge the gap between the different domains to address the complexity and heterogeneity of the NAS design. They are used to specify the semantic mapping between conflicting domains involved in NAS. Design contracts enable NAS co-design by considering constraints across domains in the control design.

### III. Timing Imperfections in NAS

NAS is a distributed system with interconnection among physical components, communication and software. Capturing the entire system in one model and analysing it is a complex task. A better alternative is to model a single component that captures all the properties of the system under study, and then to scale it. This not only simplifies the analysis, but also gives a microscopic view of the constraints in NAS. Birgit Vogel-Heuser et al. [41] presented one such model that captures the timing behaviour of typical NAS as shown in Figure. 2. It shows the usual components present in NAS and gives a comprehensive framework to study timing related problems in NAS. Plant models the physical process being controlled and the controller shown is the mechanism used for achieving the desired objective from the plant. Controller can be discrete-logic and sequencing circuit, continuous time (e.g. PID) or even hybrid (having both discrete and continuous dynamics) depending on the application. Similarly, the plant can have various dynamics and it is beyond the scope of this investigation to specify all the variations. In general, the result obtained from this investigation can be used for linear time-invariant (LTIV) systems. The sensor signals are usually sampled and send through a network which is used by the controller to generate an output to be transmitted to plant. Introduction of networks introduce imperfections in the information like latencies and data-loss which affect the timing performance of the NAS. This time-varying delay necessitates dynamic controller design methodology which can introduce undesired computation jitter [29]. Further, these imperfections also depend on the hardware used. One may infer from the above discussions that the imperfections in NAS can be captured into three categories, they are: *(i)* hardware, *(ii)* network, and *(iii)* software - induced. In order to study the timing performance of NAS it is necessary to understand the nature of these imperfections and to quantify them.

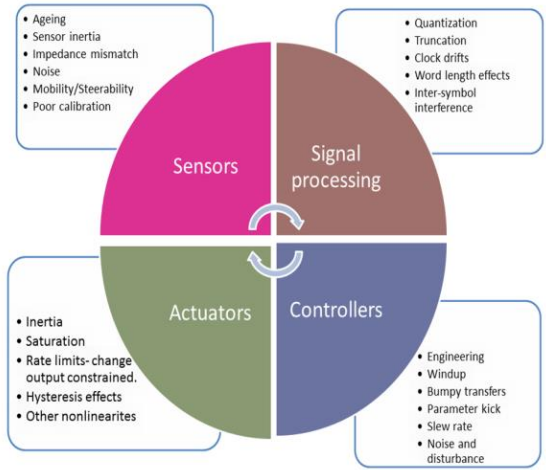

Figure 3. Source of hardware jitter in NAS from system components

Hardware timing imperfections in NAS are due to sensors, actuators, signal processing, and controller hardware as shown in Figure. 3. The effect of constraints posed by hardware components on control systems was studied in [38]. The investigation mainly studied actuator and sensor related constraints. All the constraints involved in the design of the system were not studied. Sensor induced timing imperfections have been studied extensively in the domain of instrumentation. For our analysis, we restrict our treatment to sensor drift, ageing, impedance mismatch, noise, mobility/steerability, and calibration. Signal processing related imperfections are mainly due to quantization, conversion (analog-to-digital, and digital-to-analog), truncation, clock drifts, aliasing, and inter-symbol-interference. Imperfections due to control hardware are due to windup, bumpy transfers from manual to auto, and vice-versa, slew rate of the controller (i.e. the allowable change per unit of time), noise and disturbance acting on the controller. Feedback used in controller and its design generally overcome most of these issues in the control context still some timing jitter are experience in the system behaviour. Actuator saturation [39], hysteresis [40], actuator inertia and other non-linear behaviour are the major sources of imperfections in actuators. Others studied in literature include slew-rate of the actuator which limits the amount of change per unit time.

Software induced timing imperfections can be due to scheduling, cache memory, pre-emption, interrupts, context switching, dynamic control algorithms, multiple loops and asynchronous communication between tasks. Timing imperfections in NAS require dynamic control algorithms or compensation mechanisms to ensure stability and performance [29,37], and this has been reported to lead to jitter. It is usually a practice in software engineering to measure timing using execution times (like best-case execution time, worst-case execution time, average execution time etc.) [43-48]. In our analysis, we use the concept of WCET to define the bound on the execution of control algorithm and average execution time to denote the nominal execution time. The concept of execution time leads to having a deterministic bound on the computation induced network jitter. Considering NAS to be engineered with a pre-specified hardware the concept of execution times lead to deterministic bounds.

Communication related imperfections are network latencies and data-loss. Latencies in communication channel are due to network access delays, and transmission delays. Latencies in communication channel depend on various parameters like length of the communication channel, protocol used, loading in the channel, connected nodes, real-time requirement of the connected nodes, network hardware (like network interface card), and buffers in Ethernet. Packet loss in communication channels depend on the transmission media, engineering, noise levels, and channel loading. The parameters by themselves are time-varying thereby making network latencies and loss to be random. Having studied the various sources of timing imperfections in NAS, we model them for use in specifying the contract.

## IV. MODELLING TIMING IMPERFECTIONS IN NAS

In analysis of real-time control systems the concept of jitter has been used to capture the timing imperfections in the past (see, [33-37] and the references therein). In particular sampling jitter and sampling-actuation delays have been used. The concept of time-varying sensor-to-actuator delay has been captured as delay jitter in [19,42]. In [46], it has been pointed out that the jitters not only cause performance degradation but also can destabilize the system in the worst case. Motivated by the above results, we employ the concept of jitter to denote the timing imperfection in sampling and sensor-to-actuator delays. Jitter has been identified as one of the emerging key parameter uncertainty in communication domain in [47]. There are two approaches widely available to capture jitter, they are: *(i)* Jitter measurement [48,49,54-57], and *(ii)* Jitter modeling [29,57-60]. Jitter measurement is usually suitable for hardware or small-systems. Modeling is more suitable for distributed systems as it gives generic framework for capturing the jitter occurring in various components and artifacts. Due to the distributed nature of NAS, we employ jitter modeling in our analysis.

Next step in analyzing jitter is to model it based on the occurrence based on the analysis in the previous section. Jitter has been classified based on their occurrence (behavior) as being deterministic and random in [49,58]. We use the same classification in our analysis as it suits NAS requirements. Then, we use some classifications specific to NAS like behavior, and components causing jitter as shown in Figure 4. One can visualize that hardware and software jitter are completely captured by the deterministic jitter, and communication related jitter is random. As stated earlier software induced jitter are analyzed using execution times. This means that the jitter occurring due to software can be modeled to be deterministic. Further, hardware related jitter are usually either slow varying (like ageing, drift etc.) or can be modeled to be constant. Assuming that the quantization effects have been considered in engineering of NAS, then hardware jitter can be modeled to be a bounded value.

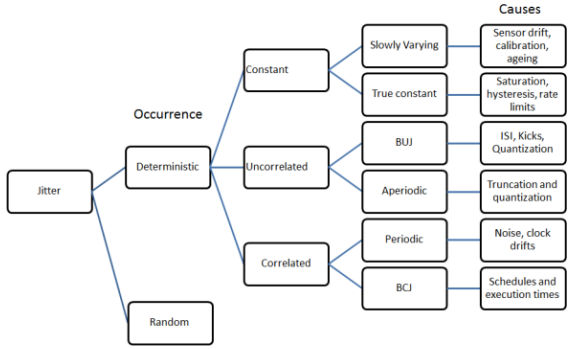

Figure 4. Source of hardware jitter in NAS from system components
(BUJ- bounded uncorrelated jitter, BCJ- bounded correlated jitter)

One can infer from the above discussion that jitter can be classified based on the component of NAS as being hardware, software, or communication-induced as shown in Figure. 5. Further, they can be modeled using their behavior as being constant, deterministic, and random. We assume that the bounded variation in the jitter due to hardware can be modeled to be a constant. This is a valid assumption as usually the jitter due to hardware change very slowly over time. Then model the software jitter to be deterministic using execution times. Finally, we capture the jitter due to network as a random phenomenon. Two models can be used to describe the random network jitter, they are: 1. empirical model, and 2. discrete-time Markov chain. Empirical model has been used in literature to model channels which exhibit stationary behavior over long time [61]. More dynamic models like MCMC have been proposed in literature in the context of NCSs [29]. DTMC model has been proposed in the context of NAS in [59], and has been used to design hybrid controller in [63].

## V. NAS CO-DESIGN

### A. TOLC contract [19]

This section is motivated to extend the concept of jitter to co-design NAS. As stated earlier, we use the concept of design contracts proposed in the context of CPS for capturing the constraints and assumption across the domains. In the analysis of real-time control systems sampling jitter and sampling-to-actuation delays has been studied in [36,37] for NCSs design. The use of sampling jitter and sampling to actuation delay in the context of CPS has been discussed in [19], and a contract called timing tolerance contract (TOLC) has been introduced. Our analysis in previous section leads to a framework for capturing the sampling jitter and sensor-to-actuator delays using jitter. The motivation of this section is to extend the TOLC contract to NAS.

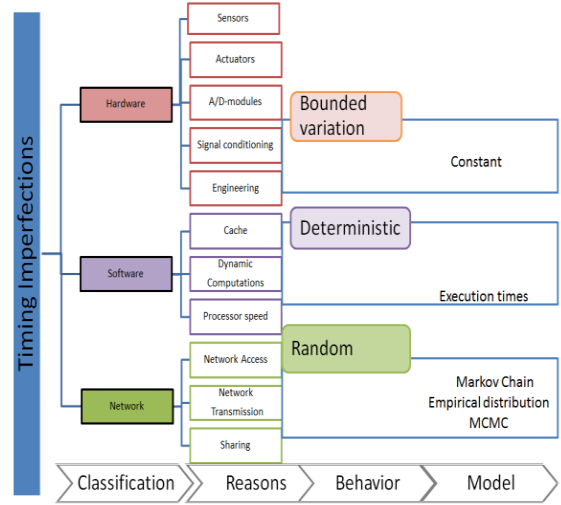

Figure 5. Timing imperfections in NAS- behavior and model

### B. TOLC implementation

Consider the single loop control system in Figure.2 which typically models the NAS. Now assume that the various jitter in the NAS has been captured using the analysis in previous section. Further, assume the network jitter is modeled using a discrete-time Markov chain wherein the states describe the loading in the channel, and the transition probability matrix is used to capture the variations in delay.

**Theorem 1** Consider NAS in Figure. 2 with the timing imperfections modeled as states of DTMC. The controller for the NAS is a hybrid controller that switches states depending on the channel conditions (i.e. loading).

This result is a direct extension of the main result in [63].

**Definition 1** A state machine M of type Mealy with extended with I/O interface functions defines the controller in Figure. 2. Such a machine M is characterized by: a set of inputs (inputs), a set of output variables (outputs), a set of state variables (state), an initialization function that initializes the state, a sampling function that reads the sensors and assigns values to the inputs, an actuation function that writes the output to the actuators, an output function that computes the output from the current input and state. The sampling and actuation are the I/O interface functions of the machine.

*Corollary 1:* Extension of theorem 1 using definition 1 in NAS shown in Figure. 2 leads to a hybrid controller of Mealy type M with the states defined by the channel loading, inputs are the delay samples, and the output is the controller mapping. A similar approach for mapping controllers using look-up-table has been proposed in [29].

**Definition 2** [19]: TOLC is specified by a tuple ($M, h, \tau, J^h, J^\tau$) where M is a state machine, $h$ and $\tau$ are the nominal period and sensor-to-actuator delay, and $J^h$ and $J^\tau$ are the bounds on the admissible variations on the period and the sensor-to-actuator delay. All parameters are assumed to be positive and satisfy the constraints $J^\tau \leq \tau$ and $J^h + \tau + J^\tau < h$, the contract states that $t_k^s \in [kh, kh + J^h]$, $t_k^a \in [t_k^s + \tau - J^\tau, t_k^s + \tau + J^\tau]$ and the k$^{th}$ state update happens before $t_{k+1}^s$ $\forall k$.

where $t_k^s$ and $t_k^a$ denote the sampling and actuation instant s respectively.

It has been pointed out in [19] that the tolerance parameter $J^h$ and $J^\tau$ need to be derived and no guidelines have been given for deriving it. Further, the procedure to model the controller as a state machine has not been proposed. Corollary 1 gives the conditions for modelling the controller as state machine of mealy type M, and one advantage of our method is it has the concept of jitter in-build in it. In order to specify the contract the jitter bounds need to be indicated. We use results from control theory to derive the jitter bounds.

**Theorem 2:** Tolerance bounds $J^h$ for the NAS in Figure. 2 with continuous-time physical process is given by

$$J^T{}_{max} < \frac{1}{|T_u(j\omega)|\omega} \qquad \forall \omega \in [0, \infty[. \tag{1}$$

*Proof:* Considering the physical process to be a continuous-time plant and the controller to be a hybrid controller that switches states depending on the sampling jitter that is modelled as the states of the Markov chain as in [63], the closed loop transfer function of the system is given by

$$T_u(s) = \frac{P(s)C_u(s)}{1 + P(s)C_u(s)} \tag{2}$$

where $C_u$ is the switching controller that switches states depending on the jitter modelled as the states of the Markov chain. To simplify our analysis, let us assume that there are two states in the Markov chain, namely- *Low* and *High*. The maximum allowable jitter margin for various loading conditions can be given in terms of the total jitter, given by $J^T = J^h + J^\tau$. Now assuming the hardware jitter to be constant, the best-case execution time (BCET) of the algorithm can be used to establish the variance in the software jitter due to software execution as

$$BCET = \sigma_s = \tau_s - J^\tau \tag{3}$$

Next, the total variance in sampling jitter can be computed from the variance of Markov chain at the different states as,

$$\sigma_T = \sigma_s + \sigma_N + \alpha_c \tag{4}$$

where $\sigma_N$ is the variance due to network, and $\alpha_c$ is the jitter due to hardware (assumed to be constant or slowly-varying). The sampling jitter distribution can be obtained with the average execution time and the average delay considering the network communication as

$$\mu_T = \mu_s + \mu_N \tag{5}$$

where $\mu_s$ is the average execution time and $\mu_N$ is the mean of the state of the Markov chain, then the sampling jitter can be modelled for different channel loadings with mean given by (5) and the variance given by (4) thereby modelling the sampling jitter in NAS. Using the result of theorem 1 in [68], we have

$$J^T{}_{max} < \frac{1}{|T_u(j\omega)|\omega} \qquad \forall \omega \in [0, \infty[.$$

This gives the total jitter margin from the sampling time to the actuation time considering the timing imperfections in NAS. One such result has been obtained in the context of wireless network control systems WNCS in [67] for control loops with communication constraints. Here, the constraints due to software and hardware have also been included in the context of NAS. From the requirement of the contract, we have

$$J^T{}_{max} \leq |h_k - h| \qquad \forall k \in \{0,1...\}$$

leading to

$$J^T{}_{max} + h \leq |h_k| \tag{7}$$

Equation (7) gives the effective sampling time $h_k$ in terms of the sampling time and maximum allowable jitter. Specifications of the design contract can be found using equations (2)-(7).

## VI. CONCLUSIONS

A co-design approach for NAS that considers control design in the presence of timing constraints posed by other competing domains in NAS has been investigated. The co-design approach uses design contract to capture the interface among the domains. Specification of design contract requires capturing the timing-imperfections in NAS, and a framework to capture them. We used the concept of timing jitter to model the various timing imperfections in NAS. Finally, we obtained results useful for the implementation of design contracts using theoretical results from control theory. Implementation of the design contract and studying it for variations physical plant (like discrete and hybrid) are the future course of this investigation.